\documentclass[epj]{svjour}
\usepackage{graphics}
\usepackage{amsmath}
\newcommand{\bnvo}{$\beta-$Na$_{0.33}$V$_{2}$O$_{5}$}
\newcommand{\bb}{K$_{0.3}$MoO$_{3}$}
\newcommand{\rb}{Rb$_{0.3}$MoO$_{3}$}
\newcommand{\tmi}{T_{MI}}
\newcommand{\mtmi}{$T_{MI}$}
\newcommand{\et}{E_T}
\newcommand{\met}{$E_T$}
\newcommand{\mtna}{$T_{Na}$}
\newcommand{\tna}{T_{Na}}
\begin{document}
\title{Nonlinear transport in \bnvo}
\author{S.~Sirbu\inst {1}\and P.H.M.~van~Loosdrecht\inst {1}\and T.Yamauchi \inst {2}\and Y.~Ueda \inst {2}}
\institute{Material Science Center, University of Groningen, 9747 AG
 Groningen, The Netherlands, \email{p.h.m.van.loosdrecht@rug.nl} \and
Institute for Solid State Physics, Tokyo University, Japan.}

\date{\today}

\abstract{Transport properties of the charge ordering compound
\bnvo\ are studied in the temperature range from 30~K to 300~K using
current driven DC conductivity experiments. It is found that below
the metal-insulator transition temperature ($\tmi = 136~K$) this
material shows a nonlinear charge density modulation behavior. The
observed conductivity is discussed in terms of a classical domain
model for charge density modulation transport.
 }
\maketitle
\section{Introduction}\label{sec:intro}
There are many solid state materials which are one - dimensional and
metallic at room temperature. Due to the coupling of the electrons
to the underlying lattice the metallic state in these materials is
usually not stable, leading to a phase transition into a charge
modulated state at low temperatures. The low temperature ground
state of the coupled electron-phonon system is characterized by a
gap in the single-particle excitation spectrum, by collective modes
formed by the electron-hole pairs, and by a deformation of the
lattice. Many of those compounds are inorganic (\bb, NbSe$_{3}$,
TaS$_{3}$) \cite{dum83,fle79,gru88}, but these properties are found
in organic compounds as well
\newline ((2.5(OCH$_{3}$)$_{2}$DCNQI)$_{2}$Li, TTF-TCNQ)
\cite{pin99,wan03}. The best known type of this class of phase
transitions is the Peierls transition \cite{pei55}. In this case the
material develops the charge density modulation state at low
temperature with a simultaneous deformation of the lattice, and the
opening of a gap in the quasiparticle excitation spectrum. As long
as the charge density modulation is pinned by lattice or impurities,
these materials can be described as narrow-band-gap semiconductors
or even insulators. An exception to this rule is NbSe$_{3}$, which
remains a semimetal at low temperature \cite{ong77}. One of the more
intriguing features of low dimensional charge ordered materials are
their nonlinear transport properties. In the charge density
modulation ground state they only exhibit ohmic conductivity below a
certain threshold applied electric field \met. Below this field, the
charge density modulation is pinned by lattice, impurities, defects,
and grain boundaries and the conductivity is solely due to a
strongly temperature dependent quasiparticle transport. For fields
above \met, the conductivity becomes strongly enhanced and nonlinear
due to the contribution of the now moving depinned charge density
modulation \cite{aya99,bra04,loo02,zaw00}. In addition to the
nonlinear behavior, charge density modulated  materials often show
an alternating current response to a static applied field. This
latter may be either due to so called ratcheting of the charge
density modulation phase as observed in for instance NbSe$_3$
\cite{fle79}, generally referred to as narrow band noise, or due to
macroscopic polarization oscillations as observed in for instance
blue bronze at low temperatures \cite{tes87}.

The recently revived interest in the vanadium bronze \bnvo\ has been
triggered by the observation of a charge ordering transition
\cite{yam99}, and sparked once more by the observation of the
pressure induced superconductivity \cite{yam02} in this
electronically low dimensional material. The vanadium bronze \bnvo\
has been the subject of various structural studies during the last
40 years \cite{sie65,yas82}. At room temperature \bnvo\ is a highly
anisotropic metal with a site occupancy disorder of the sodium
atoms. Around $T_{Na}$ = 240~K a second order phase transition
occurs leading to an ordering of the sodium atoms and a doubling of
the primitive cell along the b direction \cite{ued01}. A charge
ordering transition, which is thought to be driven by the electron
phonon coupling, occurs at $\tmi$ = 136~K and is accompanied by a
further tripling of the unit cell along the b direction
\cite{ued01,nag05}, leading to a commensurate charge modulated state
with a period of 6b. Temperature dependent measurements of the
magnetic susceptibility revealed a magnetic transition from a
paramagnetic to an antiferromagnetic state at $T_{AF}=22$~K
\cite{yam99}. Finally, as mentioned above, \bnvo\ becomes a
superconductor for pressures above 8~GPa and low temperatures (8~K)
as revealed by recent pressure dependent resistivity measurements
\cite{yam02}. Optical conductivity data suggest that this system may
be understood in terms of a small polaron model, where the charge
ordering in fact corresponds to an ordering of the polarons
\cite{pre03,pres03}. Since previous experiments strongly suggest
that \bnvo\ is a low dimensional conductor with an electron-phonon
interaction induced charge ordering, this material should also show
the usual nonlinear transport properties discussed above. Indeed the
results presented in this paper of the first detailed measurements
of the field dependence of conduction in the sodium bronze \bnvo\
are fully consistent with this picture. The observed nonlinear
transport properties are well described using a classical domain
model. The observed charge density modulation conductivity increases
with increasing temperature, suggesting a screening of the charge
density modulation pinning by the thermally excited carriers.
\section{Temperature dependent transport}\label{sec:tdep}
Single crystal samples have been prepared as described elsewhere
\cite{ued01}. Platelets with typical dimensions $4$ $\times$ $2$
$\times$ $0.2$ mm $^{3}$ were mounted on the cold finger of a flow
cryostat and contacted using $50$~$\mu$m diameter platinum wires.
Four wires were fixed on the sample surface using silver paste,
spaced 1 mm apart. Measurements were performed along the b axis in a
four probe configuration using a Keithley 236 source-meter in a
current driven mode. The resistivity was calculated from the
measured resistance and the geometry of the sample.

\begin{figure}
\resizebox{0.5\textwidth}{!}{%
  \includegraphics{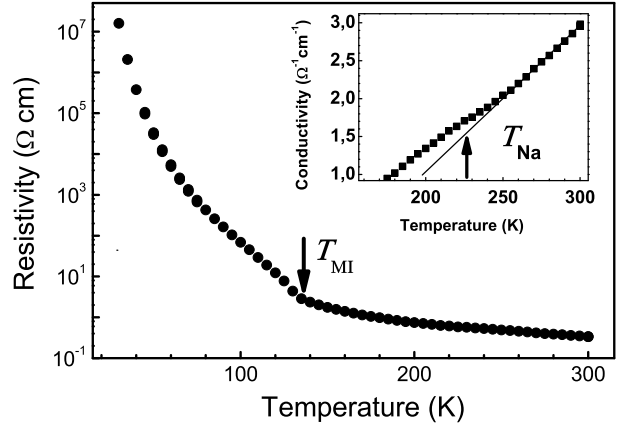}
}
\caption{  Resistivity as a function of temperature in
  \bnvo. Inset: change in conductivity near the
  sodium ordering transition at $\tna\simeq 240$~K.}
\label{fig:tdep}
\end{figure}

 First we turn to the
temperature dependence of the low-field ohmic resistance, of which a
typical example is shown in Fig.\ref{fig:tdep} \cite{samples}. The
strong increase of the resistivity below $T_{MI}$ = 136~K clearly
signals the charge ordering transition, consistent with results
published in literature. Note that also the sodium ordering
transition at $\tna = 240$~K leads to a change in the temperature
dependence of the resistivity (see also inset Fig. \ref{fig:tdep}.).
The small enhancement of the conductivity below \mtna\ can be
understood in terms of the decreased amplitude of the spatial
potential variations on the vanadium sites due to the ordering of
the sodium subsystem \cite{pre03}. Although the resistivity at high
temperatures ($T>\tmi$) is fairly small, as expected for a bad
metal, it does not exhibit the metallic behavior as observed
previously \cite{ued01}. The reason for this could be that the
samples are slightly misaligned. In this case the measured
resistivity is a combination of the resistivity of the metallic
b-axis and the insulating perpendicular axes, leading to the
observed temperature dependence. X-ray diffraction experiments on
the samples, however, have ruled out this option in showing that to
within a degree the samples are indeed b-oriented. A more likely
reason is that the sodium stoichiometry of the sample slightly
deviates from $x = 0.33$, which is known to lead to a rapid loss of
metallic behavior and eventually to the disappearance of the
metal-insulator transition \cite{yam99}. Such deviations lead to an
additional disorder in the sodium site occupancy, which in turn
leads to a more disordered potential on the vanadium sites which
make up the one-dimensional conduction chains \cite{pre03}. Since in
particular low dimensional systems are very susceptible to disorder,
this may lead to a small disorder induced gap and to the observed
non-metallic behavior.

An estimate for the transport gaps above and below the charge
ordering transition temperature is obtained fitting a simple
activated behavior, $\rho = \rho_{0}\exp(\Delta/kT)$, to the data.
Well within the charge ordered phase (30~K~$\leq T\leq$~80~K) we
find $\rho_0=630$~m$\Omega${cm} and a gap $\Delta_{LT} = 548$~K. At
high temperatures (150~K~$\leq T\leq$~300~K) we find
$\rho_0=75$~m$\Omega${cm} and a gap $\Delta_{HT} = 472$~K.
Surprisingly, the transport gap in the charge ordered phase is only
about 15\% larger than the one found for the high temperature phase.
The major change is found in the prefactors which differ by an order
of magnitude. This is consistent with the expectation that the
majority of the charge carriers is frozen out by the charge ordering
at \mtmi. In a similar analysis Yamada {\em et al.} \cite{yam99}
found an activation energy of 538~K, in good agreement with the
present value. In previous work we found an optical gap of
$2\Delta$=2450~K \cite{pre03}. Most likely, this optical gap
corresponds to the energy needed to free a quasiparticle from the
charge ordered state. Clearly then, the transport gap must have a
different origin. The origin for this difference could be that the
observed transport gap is in fact a sodium disorder induced pseudo
gap in the charge carrier spectrum, where the charge carriers
themselves are present due to an incomplete charge ordering, rather
then due to thermal excitation as is usually the case. This would be
consistent with the small difference in the transport gap values
below and above the charge ordering transition, as well as with the
observed finite low frequency conductivity in the optical data
\cite{pre03}.

\begin{figure}[hbt]
\resizebox{0.5\textwidth}{!}{%
  \includegraphics{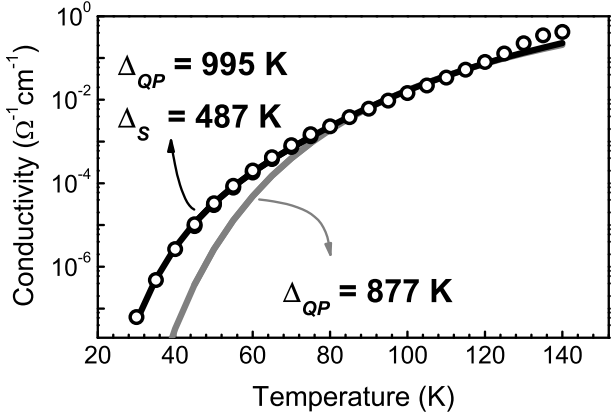}
} \caption{ Temperature dependent conductivity (symbols, same data
as in Fig. 1.)
 together with a fit to an activated two particle behavior, Eq. (1) (dark line).
 The gray line shows a fit of thermally activated quasiparticle transport to the high
 temperature part of the data.} \label{fig:sol}
\end{figure}

However, closer inspection of the low temperature conductivity shows
that it does not strictly follow an activated behavior. In
particular, below 60~K there seems to be an enhancement of the
conductivity, which presumably is due to the presence of an
additional conductivity channel. In a purely one-dimensional model,
such an enhancement may originate from the presence of mid-gap bound
states of amplitude $\pi$-solitons and quasiparticles, as described
by Brazovskii \cite{bra80}. Indeed, the low temperature conductivity
is better described by the empirical from
\begin{equation}\label{eq:actsol}
 \sigma (T)= \sigma^{0}_{QP}\exp(-\Delta_{QP}/kT)+\sigma^{0}_{S}\exp(-\Delta_{S}/kT).
\end{equation}
Here the first term on the right hand side accounts for the
contribution of the quasiparticles, whereas the second term takes
the midgap state conductivity into account. In the {\em
one}-dimensional description, the midgap states are located halfway
the quasiparticle gap, {\em i.e.} $\Delta_{S}\approx \Delta_{QP}/2$.
Fitting Eq.~(\ref{eq:actsol}) to the data yields $\Delta_{QP}$ =
995~K; $\Delta_{S}$ = 487~K, and a ratio
$\sigma^{0}_{QP}/\sigma^{0}_{S}\sim 0.4\times10^3$. Note that the
quasiparticle gap obtained in this way becomes comparable to the gap
found in optical experiments \cite{pre03}. The conductivity ratio
shows that just below the phase transition quasiparticle transport
dominates the conductivity, whereas the 'midgap' contribution
becomes important at lower temperatures only, despite its smaller
energy gap. Finally, we note that a similar analysis to blue bronze,
\bb, data leads to similar conclusions. For both these cases, one
can worry whether a one-dimensional model is applicable at all.
Indeed, the midgap states described by Brazovski require the
existence of topological $\pi$-solitons which do not exist in three
dimensions. Therefore, the above analysis merely shows the presence
of additional excitations halfway the gap. Though in sodium
vanadate, these might originate from the sodium disorder, one would
not expect a similar disorder in blue bronze. Therefore, the precise
nature of the midgap states remains unresolved at present. One
interesting thought is that they might result from excitations
inside domain walls which separate the charge density modulation
ordered regions in the samples.
\section{Field dependent transport}\label{sec:edep}
The charge density modulation in low dimensional systems is usually
pinned by the underlying lattice, by impurities or by structural
defects. When the charge density modulation is incommensurate with
the underlying lattice, the main pinning centers are the impurities
and other defects. In this case, the pinning might be considered
relatively weak, and the charge density modulation can fairly easily
move in the material upon application of an external electric field,
leading to the typical non-linear charge density wave conduction.
When the modulation is commensurate with the lattice, the pinning is
considered to be strong, usually much stronger than the impurity
pinning, and the large pinning energy prevents conduction of the
charge density modulation. The charge modulation in \bnvo\ is
commensurate with the underlying lattice. Since the modulation
period is, however, rather large (6b) the commensurability pinning
in this system is expected to be relatively weak, opening the
possibility of non-linear charge density wave conduction in the
charge ordered phase.

\begin{figure}[hbt]
\resizebox{0.5\textwidth}{!}{%
  \includegraphics{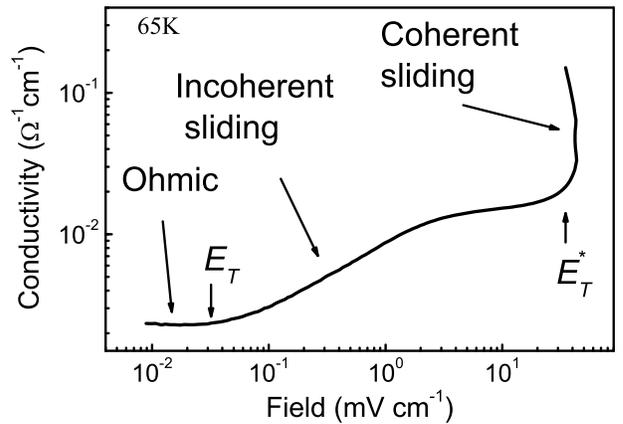}
} \caption{Nonlinear field dependent conductivity in \bnvo\ measured
along the b axis at 65~K displaying
 three charge density modulation transport regimes. } \label{fig:nlc}
\end{figure}

To study the nonlinear transport properties of single crystal \bnvo,
current driven field dependence measurements \cite{field} were
performed in the temperature range 65~K - 300~K. The obtained
transport properties are very similar with those obtained in CDW
systems. A typical example of the conductivity measured along the
b-axis at 65~K is displayed in Fig. \ref{fig:nlc}.

The behavior of the conductivity shows three regimes. Below
0.06~mV/cm (first threshold field) the conductivity is
field-independent and mainly due to quasiparticle transport. At
about 0.06~mV/cm the conductivity shows a nonlinear increase
resulting from an incoherent contribution of the charge density wave
to the conductivity. This behavior saturates around 4~mV/cm. Above
the second threshold field at 30~mV/cm a steep increase of the
conductivity takes place signaling the onset of the coherent charge
density modulation regime. Despite the strong increase of the
conductivity, it never reaches values comparable to the metallic
state.

The values for the threshold fields are smaller than those found,
for example in blue bronze. Zawilski {\em et al.} \cite{zaw00}
reported 40~mV/cm for the first threshold field and around
2000~mV/cm for the second threshold field in \bb\ at 60~K while
Mihaly {\em et al.} \cite{mih88} found 40~mV/cm in
K$_{0.3-x}$Na$_{x}$MoO$_{3}$ around the same temperature. Fleming
{\em et al.} \cite{fle86} found the first threshold field around
500~mV/cm in TaS$_{3}$ and 90~mV/cm in \bb\ both measured at 60~K.
In K$_{0.3-x}$Na$_{x}$MoO$_{3}$ (x = 0, 0.02, 0.05, 0.1), at 77~K,
Wang {\em et al.} \cite{wan99} reported values of (300, 670, 750,
950) mV/cm for the first threshold field at 180~K. Beauchene
\emph{et. all} \cite{bea86} reported similar values in \rb.
K\"{u}ntscher {\em et al.} \cite{kun05} found in \bb\ and \rb\
values of 150~mV/cm for the first threshold field at 60~K while van
Loosdrecht {\em et al.} \cite{loo02} found 200~mV/cm at the same
temperature. Although the values of the first threshold field for
\bb\ show some variation, they are typically in the 50-200~mV/cm
range, $i.e.$ substantially higher than the present values observed
for \bnvo. This is consistent with the notion that the temperature
dependence of the low field transport observed in \bnvo\ is due to
incomplete charge ordering resulting from disorder in the sodium
sublattice. The presence of charge carriers, even at low
temperatures, leads to screening of the pinning potential, and hence
to a lowering of the threshold field. In contrast, the presence of
charge carriers in blue bronze is almost entirely due to thermal
quasiparticle excitations.

\begin{figure}[hbt]
\resizebox{0.5\textwidth}{!}{%
  \includegraphics{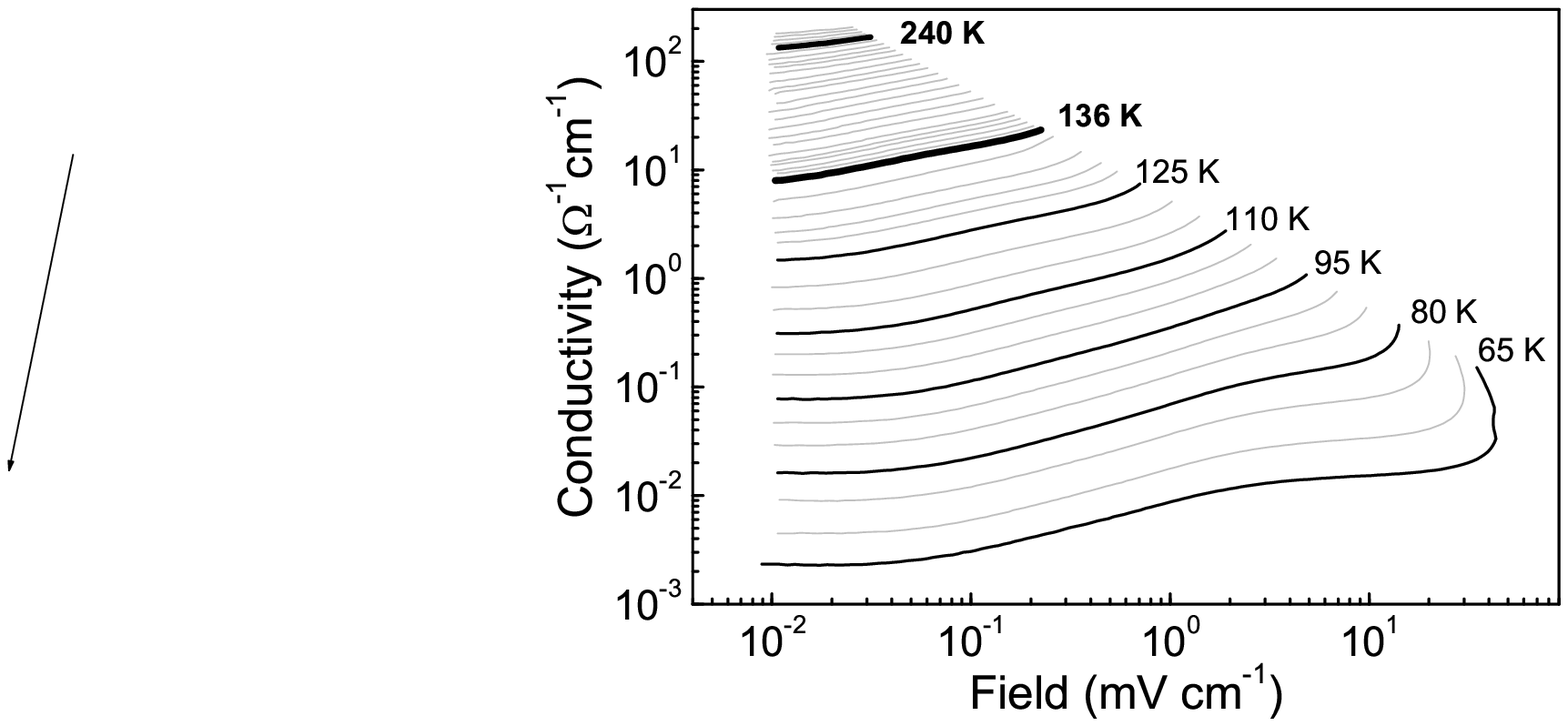}
} \caption{ Nonlinear conductivity along the b axis as a function of
electric field in
  \bnvo\ measured from $65$~K to $300$~K
  (in $5$~K steps between 65~K - 125~K, 140~K - 230~K, and 250~K - 300~K;
  in 2~K steps in the transition temperatures regions: 128~K - 146~K, and 236~K - 246~K).} \label{fig:nlct}
\end{figure}
Figure \ref{fig:nlct} displays the electric field dependence of the
conductivity along the b-axis for a variety of temperatures between
65~K and 300~K. The nonlinear behavior is most pronounced at low
temperatures, and is slowly decreasing upon increasing the
temperature towards the transition temperature $\tmi$. The first
threshold field $\et$, {\it i.e.} the field required to induce
charge density modulation conductivity, is observed below $\tmi$
only. As the temperature is reduced, the threshold fields increase,
evidencing a strengthening of the charge density modulation pinning.
This is due to the reduction of the free carrier concentration at
lower temperatures, leading to a less effective screening of pinning
centers. The second transport regime, the incoherent moving regime,
shows an increasing conductivity followed by saturation upon
increasing field. Going up in temperature, the field at which this
saturation is reached
increases, until around 90~K it merges with the second threshold field.\\
Even above the charge ordering transition temperature, the nonlinear behavior has not
entirely disappeared, although there are no clear threshold fields anymore.
Nonlinear
conduction is still observed up to the sodium ordering temperature
$\tna$ = 240~K, consistent with the disorder induced non-metallic
behavior observed above $\tmi$. Finally, above 240~K a nearly
field independent conductivity is observed.

\begin{figure}[hbt]
\resizebox{0.5\textwidth}{!}{%
  \includegraphics{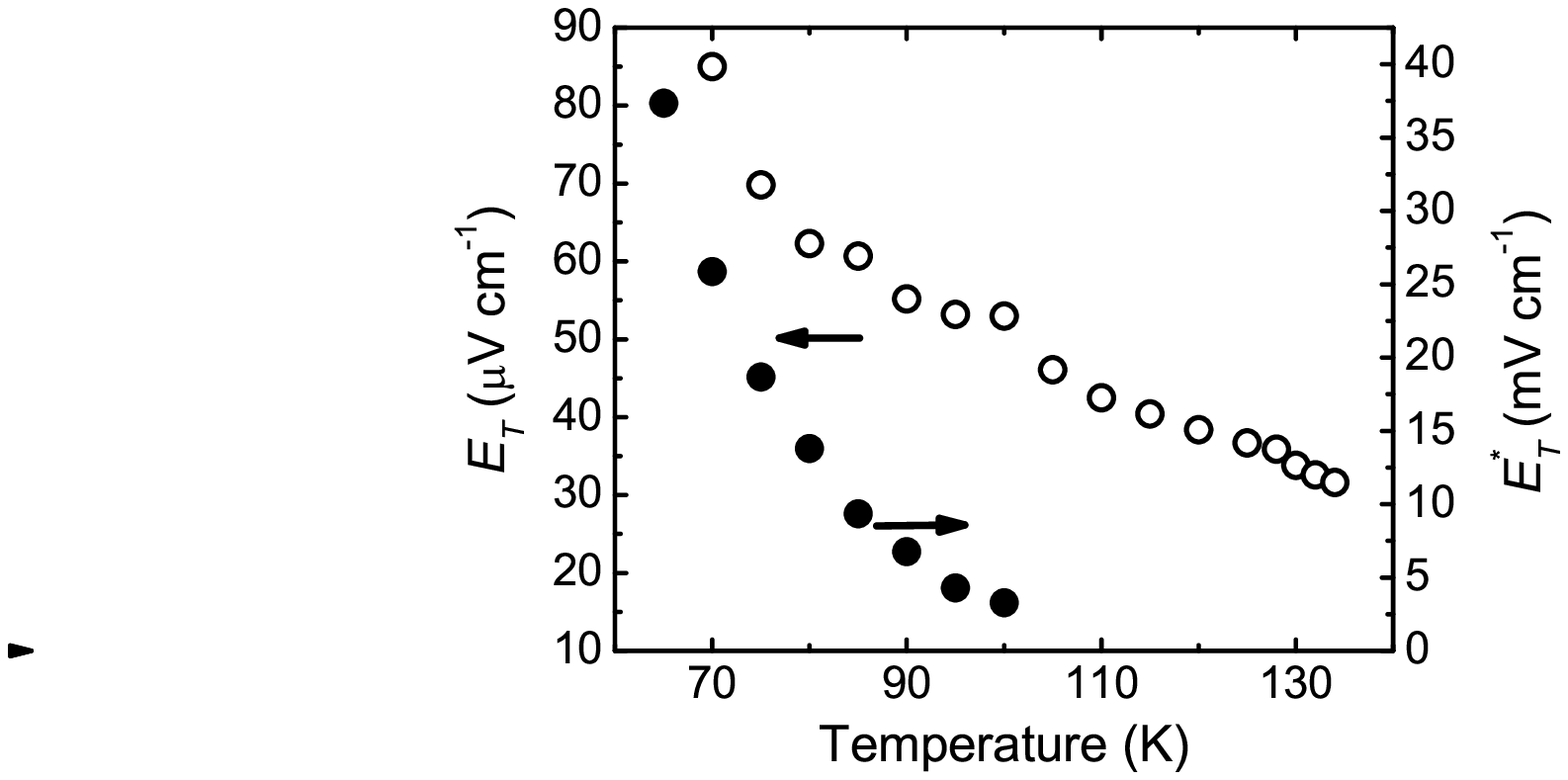}
} \caption{Temperature dependence of the threshold fields for
$T<\tmi$, obtained
   from the conductivity data. Open symbols (left scale): first threshold field;
   Filled symbols (right scale): second threshold field. } \label{fig:tres}
\end{figure}
The temperature dependencies of the first and second threshold
fields are shown in Fig. \ref{fig:tres}. The open symbols display
the first threshold field, $\et$, obtained from the conductivity
curves by taking the values where the conductivity starts to
increase. The lower part displays the second threshold field,
$E_{T}^{*}$, obtained in a similar manner. Although there is a
factor of 10$^3$ difference between the two threshold fields, the
displayed behavior is qualitatively the same: the threshold field
strongly decreases with increasing temperature. With some exceptions
( for example \bb\ or TaS$_{3}$ \cite{sch89} ) this type of behavior
for in particular the first threshold field is common for many of
the known charge density wave systems. At low temperature, when the
number of quasiparticles is reduced, large electric fields may build
up around pinning centers and the value of $E_{T}^{*}$ is
essentially determined by the amount of disorder in the system .
Going up in temperature, the thermally excited quasiparticles tend
to homogenize the electric field inside the sample. This effect is a
natural source for the redistribution of the driving fields inside
the sample.

A variety of models were proposed to describe the nonlinear
conductivity observed in charge density modulation materials, mostly
based on the original suggestion of Fr\"{o}hlich \cite{fro54} that
conductivity is dominated by sliding CDW transport. Between them,
there are two main approaches in describing the CDW conductivity
\cite{gru88,bra04,oga05}. The first one, treats the CDW as a
classical particle which is moving in a periodic potential, with the
period determined by the period of the CDW \cite{mon82,gru81}. This
model gives a sharp threshold field, $E_{T}$ for the onset of
nonlinearity and a saturation of the conductivity for high electric
fields. The second model, proposed by Bardeen and referred to as the
tunneling model \cite{bar79}, assumes that the nonlinear transport
occurs as a result of coherent tunneling of the CDW over macroscopic
distances. Besides the threshold field, $E_{T}$, the model gives
another characteristic field $E_{0}$, which can be interpreted as a
tunneling barrier. The onset of the conductivity reveals a sharp
threshold field and at high electric fields the conductivity
saturates. Both models rely on a $T = 0$ treatment of the problem,
though nonzero temperature models based on thermally assisted flux
creep have been discussed as well \cite{lem99,oga05}. The models
described above fail to account for the charge density modulation
conductivity observed in \bnvo. Fukuyama, Lee and Rice
\cite{fuk78,lee79} discussed the effect of the pinning centers on
the charge-density-waves dynamics. They focused on phase
fluctuations and show that, when the material is characterized by
weak pinning, the system can be though to break up into domains.
Here, we integrate this notion in a simple phenomenological model
describing the observed nonlinear transport in \bnvo, by treating
the sample as a collection of interconnected domains. Each domain is
characterized by its own conductivity and threshold field so that
the sample can be considered as a collection of parallel and series
nonlinear conduction paths. At low applied fields (smaller then
$\et$), excited quasiparticles dominate the conductivity of a single
domain, leading to the strongly temperature dependent ohmic
behavior. All charge density modulation domains are pinned by
lattice and defects and will not start moving until the applied
field exceeds a certain critical field. At higher applied fields,
some domains start to become depinned, leading to an additional
charge density modulation contribution to the conductivity. The
charge density modulation  still cannot move as a whole due to the
domain structure resulting from grain boundaries and strong pinning
centers. The sample is now in the incoherent moving regime. Finally
for applied fields exceeding a second critical field, the charge
density modulation may move as a whole (or at least a percolation
path exists between the contacts), and the conductivity becomes
completely dominated by the moving charge density modulation
transport. Within this model, the second critical field is then a
percolation threshold and the transport regime above this field is
dubbed coherent moving regime.

The model sketched above can be made more quantitative by specifying
a model for the charge density modulation conductivity itself. Here
we take one of the simplest approaches, and describe the charge
density modulation within a domain as a charged particle moving due
to an applied field $E$ in a viscous medium \cite{gru88}. The charge
density modulation conductivity is then given by.
\begin{equation}\label{eq:bare}
\sigma_{cdm}(E) = {\sigma_{c}\ \frac{E-E_{T}}{E}\ \theta(E-E_{T})}\
,
\end{equation}
where the Heaviside function $\theta(E-E_{T})$ assures that the
equation is also valid for $E<E_T$. In the incoherent moving regime
the conductivity can then be modeled as a collection of
interconnected charge density modulation domains, shunted by the
free carrier conductivity. In general, this is still a complex
system which solution would require detailed knowledge of domain
properties and their connectivity. Here we will take a more
qualitative and statistical approach, and model the charge density
modulation system as a statistically large number of interconnected
domains. The conductivity of this network of domains together with
the free carrier conductivity will then yield the total
conductivity. The network can be considered as a collection of
parallel and series conduction paths. For the series connected
domains of a single conduction path, the conductivity will have the
same form as Eq. (\ref{eq:bare}), but with an effective threshold
field $\hat{E}_T=\sum_i{E_T^i}$ and effective conductivity
$\hat{\sigma}_c=\left(\sum_i{(\sigma_c^i)^{-1}}\right)^{-1}$. The
total charge density modulation conductivity is then given by
\begin{equation}\label{eq:full}
\sigma_{cdm}(E) = \sum_{j}{\hat{\sigma}_{c}^{j}\
\frac{E-\hat{E}_{T}^{j}}{E}\ \theta(E-\hat{E}_{T}^{j})}\ ,
\end{equation}
where $j$ runs over the parallel conduction pathways, each having
their own effective threshold field $\hat{E}_{T}^{j}$ and effective
conductivity $\hat{\sigma}_{c}^{j}$. If there are a large number
of conduction paths, the above summation can be replaced by an
integration, from zero to infinity, weighted by the statistical
distribution of threshold fields and conductivities. For simplicity,
we assume a lorentzian distribution of width $\gamma$,
centered on $E_T^0$ for
the threshold fields, and a constant effective conductivity
$\sigma_{c}^{j}=\hat{\sigma}_c$ for each path. The total conductivity,
including the free carrier contribution, $\sigma_{0}$, can then
be equated as
\begin{equation}\label{eq:shunt}
 \sigma(\epsilon) =
        \sigma_{0}+\hat{\sigma}_{c}\frac{2\epsilon(\arctan(\epsilon)+\arctan(\epsilon_0))-\ln(
              \frac{1+\epsilon^{2}}{1+\epsilon_0^{2}})}{(\epsilon+\epsilon_0)(\pi+2\arctan(\epsilon_0))}
          ,
\end{equation}

with $\epsilon_0=\hat{E}_T^0/\gamma$,
$\epsilon=(E-\hat{E}_T^0)/\gamma$. In the limit
$\gamma/\hat{E}_T^0$ $\gg$ 1, which is valid for the present
experiment, the total conductivity becomes:

\begin{equation}\label{eq:sim}
 \sigma(\epsilon) =
        \sigma_{0}+\hat{\sigma}_{c}\frac{2\epsilon\arctan(\epsilon)-\ln(
              1+\epsilon^{2})}{\epsilon\pi}
          ,
\end{equation}

where $\epsilon=E/\gamma$

In the derivation of Eq. (\ref{eq:sim}) quite a few assumptions have
been made. The most important one is that domains are formed in the
charge density modulation material, within which the charge density
modulation transport is coherent. This is mainly based on the
observation of the slow onset of the moving conductivity in
$\beta-$Na$_{0.33}$V$_{2}$O$_{5}$ (see Fig. \ref{fig:nlc}. and Fig.
\ref{fig:nlct}.) as well as in for instance \bb\ \cite{loo02}. The
assumption of parallel paths of series resistors is pretty robust;
allowing for connectivity between these paths would only lead to
additional parallel pathways for conduction. Of course, there is
nothing known on the statistical distribution of domain properties.
Taking a symmetric lorentzian (or gaussian) distribution simply
makes the integration over domains tractable. Taking an asymmetric
distribution would be more physical. Also typical domain sizes are
not known. X-ray diffraction measurements \cite{yama02}, however,
have shown that the width of the superstructure peaks originating
from charge ordering is comparable to the sharpness of the
fundamental peaks, thereby setting a lower limit of the domain sizes
to $\sim$100 nm. Finally, we did not allow for a distribution of
effective conductivities for the conduction paths. Numerical
simulations taking lorentzian distributions $\hat{\sigma}^{j}_{c}$\
have shown, however, that the shape of the field dependent
conductivity does not strongly depend on this.

\begin{figure}[hbt]
\resizebox{0.5\textwidth}{!}{%
  \includegraphics{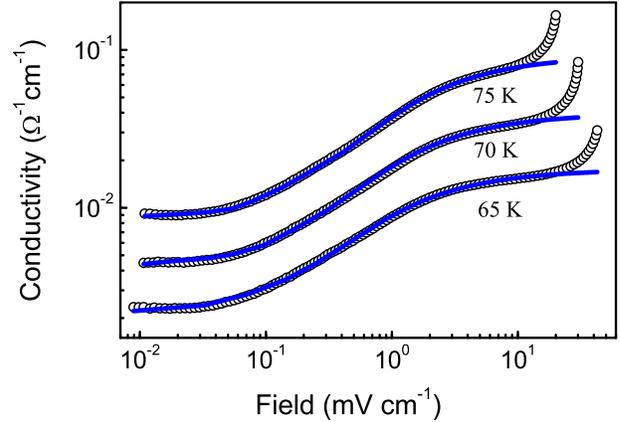}
} \caption{Fits  of the domain model, Eq. (\ref{eq:sim}) (solid
lines), to the nonlinear transport data (open symbols)
   at 65~K, 70~K and 75~K.} \label{fig:fit1}
\end{figure}

Now return to the data in Fig. \ref{fig:nlct}. We have fitted Eq.
(\ref{eq:sim}) to the field dependent conductivities, measured at
different temperatures between 65~K and 136~K. Note that the only
free parameters in the fits are the width of the distribution
$\gamma$, and the charge density modulation conductivity
$\hat{\sigma}_{c}$, since the normal conductivity, $\sigma_{0}$, can
be obtained directly from the low field data. At low temperatures, a
good approximation of $\hat{\sigma}_{c}$ can be made using the field
dependent conductivity data taking the conductivity difference
between the ohmic regime and the saturation regime. The only
remaining fit parameter in this situation is the width of the
lorentzian distribution, $\gamma$. The fits generally show a good
agreement with the data, and some typical results of the fitting are
shown in Fig. \ref{fig:fit1}.

\begin{figure}[hbt]
\resizebox{0.5\textwidth}{!}{%
  \includegraphics{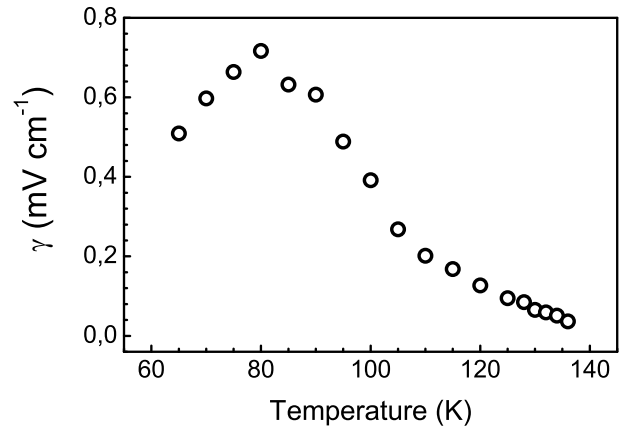}
} \caption{ Temperature dependence of the width $\gamma$ of the
lorentzian threshold field distribution
  obtained from fits of the data in Fig. \ref{fig:nlct}. to the domain model, Eq. (\ref{eq:sim}).} \label{fig:gam}
\end{figure}

The temperature dependence of the threshold field distribution
width, $\gamma$, obtained from the fits is displayed in Fig.
\ref{fig:gam}.

At low temperature, the width increases with increasing temperature.
It shows a pronounced peak at 80~K, after which it starts to
decrease becoming very small in the region of the MI transition. The
decrease with increasing temperature observed above 80~K is what one
might intuitively expect; the enhanced screening will decrease the
typical domain threshold fields, thereby decreasing $\gamma$. Apart
from the decrease of the average pinning potential, there will also
be changes in the statistical distribution as the temperature is
lowered. At low temperatures one expects that there will be a larger
number of domains with a relatively small pinning potential,
reducing the distribution width. Therefore we believe that the
increase of the width observed in the low temperature results from a
competition of enhanced screening and the formation of larger, more
strongly pinned, domains due to the coalescence of small, weakly
pinned, domains as the temperature increases.

\begin{figure}[hbt]
\resizebox{0.5\textwidth}{!}{%
  \includegraphics{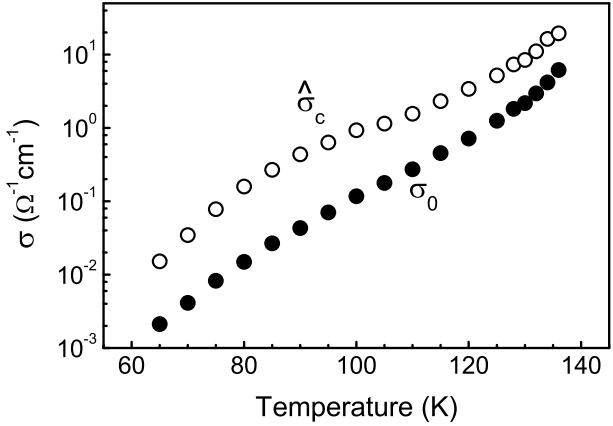}
} \caption{ Temperature dependence of the effective charge density
     modulation conductivity, $\hat{\sigma_{c}}$ (open symbols), and the low field
     ohmic conductivity $\sigma_0$ (filled symbols),
     obtained from fits of the data in Fig. \ref{fig:nlct}.
     to the domain model, Eq. (\ref{eq:sim}) (see text).} \label{fig:con}
\end{figure}

Figure \ref{fig:con} shows the temperature dependence of the
effective charge density modulation conductivity,
$\hat{\sigma}_{c}$, and the low field ohmic conductivity,
$\sigma_0$. The moving charge density modulation conductivity is
found to be almost an order of magnitude larger than the
quasiparticle contribution. The small kink in the charge density
modulation contribution around 90 K, which probably again results
from the competition between screening and coalescence.

The temperature dependence of both contributions show a similar
activated behavior. For the quasiparticles this has already been
discussed in section \ref{sec:tdep}. The usual interpretation of the
activated behavior of the moving density modulation contribution is
in terms of thermally activated flux creep \cite{lem99}, similar to
the flux creep of the Abrikosov flux lattice in superconductors
\cite{and64}. From fits to an activated behavior we estimate the low
temperature ($T<90$~K) activation energies for the quasiparticle and
the moving density modulation contributions to be 740~K and 805~K,
respectively. For the ohmic contribution, this is in good agreement
with the earlier results (Sec. \ref{sec:tdep}).

\begin{figure}[hbt]
\resizebox{0.5\textwidth}{!}{%
  \includegraphics{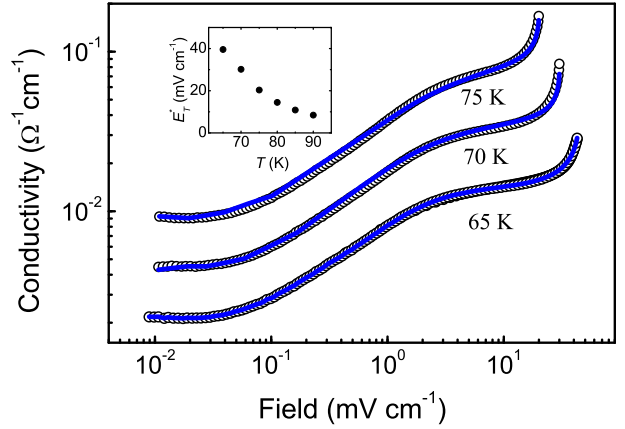}
} \caption{ Fits of the domain-percolation model, Eq.
(\ref{eq:full}) (solid lines)
   to the nonlinear transport data (open symbols)
   at 65~K, 70~K and 75~K. The inset shows the behavior of the percolation
   threshold with temperature.} \label{fig:fit2}
\end{figure}

The above discussed model gives a phenomenological understanding of
the ohmic and incoherent transport regimes. It does not, however,
describe the coherent transport regime, where a sharp rise in
conductivity takes place. The appearance of this second threshold
field can be understood as follows. As the field increases, more and
more local pinning potentials will be overcome, increasing the
number of domains which contribute to the conductivity. Eventually
this will lead to the formation of a percolation path between the
contacts. We thus propose that the second threshold is in fact a
percolation threshold. For fields bigger than the second threshold
field, $E_{T}^{*}$, the CDW will then coherently move along such
percolation paths, leading to the sharp rise in conductivity. This
leads to an additional contribution to the conductivity of the form
of Eq. (\ref{eq:shunt}), so that the total conductivity now becomes

\begin{equation}\label{eq:all}
\begin{split}
\vspace*{5cm}
 \sigma(\epsilon) =
        \sigma_{0}+\hat{\sigma}_{c}\frac{2\epsilon\arctan(\epsilon)-\ln(
             1+\epsilon^{2})}{\epsilon\pi}+ \\
             \hat{\sigma}_{p}\frac{2\epsilon_{p}(\arctan(\epsilon_{p})+\arctan(\epsilon_{0p}))-ln(
              \frac{1+\epsilon_{p}^{2}}{1+\epsilon_{0p}^{2}})}{(\epsilon_{p}+\epsilon_{0p})
              (\pi+2\arctan(\epsilon_{0c}))}
              \end{split}
\end{equation}

where $\epsilon_{p} = (E-E_{T}^{*})/\gamma_{p}$, and $\epsilon_{0p}
= E_{T}^{*}/\gamma_{p}$. $\gamma_{p}$ and $\hat{\sigma}_{p}$ are the
width of the distribution and the percolation charge density
modulation conductivity, respectively. We have fitted this last
equation to the low temperature ($T<90$~K) field dependent
conductivities and some typical results of the fitting are shown in
Fig. \ref{fig:fit2}.

Clearly the data follows Eq. (\ref{eq:all}) quite well. The
temperature dependence of the upper threshold field (see inset Fig.
\ref{fig:fit2}.) closely follows the results presented in Fig.
\ref{fig:tres}. The reason for the sharp decrease of the percolation
threshold field upon increasing temperature is the same as for the
incoherent moving threshold, as has been discussed in section
\ref{sec:tdep}, namely the increased screening of impurities upon
increasing temperature. Finally, we note that the fitting is fairly
insensitive to the distribution width as long as
$\gamma_p<<E_T^{\ast}$, as expected for a percolation threshold.

\section{Conclusions}\label{sec:con}

We presented detailed nonlinear transport experiments on \bnvo\ in
the temperature range 30~K - 300~K. The low field data in the charge
ordered phase show that two types of excitations contribute to the
transport. The charge density modulation quasiparticle gap is found
to be about 700-800 K, and likely depends on the sodium
stoichiometry. Evidence for a second type of excitation, with a gap
of $\sim 500$~K, has been presented, although the exact origin of
this excitation remains unclear at present. It might be either a
bound state of collective charge density modulation excitations like
the phason, or possibly an excitation within the domain walls
between ordered charge density modulation domains. A competing model
for the thermally activated low field charge transport in a low
dimensional system in the presence of disorder is the variable range
hopping (VRH) model \cite{mot69,mot79}. Analyzing the low
temperature data in terms of VRH conductivity indeed leads to a
reasonable agreement with this model as well, again with deviations
at temperatures below 50 K. Since the present data can not
distinguish between these models we adapted the most widely used
model here as well.

The field dependent data clearly show the charge density modulation
nature of the insulating phase, and is very similar to the transport
in other well known charge density modulation materials like \bb.
The phenomenological domain model for nonlinear transport in charge
density modulation materials presented here is found to be in good
agreement with the experiments, and we believe that this model
should be applicable to other semi-conducting charge ordered
materials as well. The model could be improved by taking a more
realistic model for the statistical distribution of domain
properties, the single domain transport, and by allowing for field
dependent domain properties. In particular the pinning fields are
expected to be field dependent, since at high moving velocities the
charge density modulation excitations may excite quasiparticle
excitations resulting from charge density friction. Finally, we
believe that in low free carrier density ({\em i.e.} semiconducting)
compounds like \bnvo\ and \bb\ much of the temperature dependence of
the transport properties, including the observed decrease of
threshold fields upon raising temperature, can be understood in
terms of enhanced screening of pinning centers by thermally excited
charge carriers.

Acknowledgements --
This project is supported by the Netherlands Foundation for
Fundamental Research on Matter with financial aid from the
Nederlandse Organisatie voor Wetenschappelijk Onderzoek.
\clearpage

\begin{thebibliography}{bnvo}

\bibitem{gru88}
G. Gr\"{u}ner, Rev. Mod. Phys. \textbf{60}, 1129 (1988).

\bibitem{dum83}
J. Dumas, C. Schlenker, J. Marcus and R. Buder, Phys. Rev. Lett.
\textbf{50}, 757 (1983).

\bibitem{fle79}
R. M. Fleming and C. C. Grimes, Phys. Rev. Lett. \textbf{42}, 1423
(1979).

\bibitem{pin99}
Marko Pinteric, N. Biskup and Silvia Tomic and J. U. von Schutz,
Synth. Metals \textbf{103}, 2185 (1999).

\bibitem{wan03}
Z. Z. Wang, J. C. Girard, C. Pasquier, D. Jerome and K. Bechgaard,
Phys. Rev. B. \textbf{67}, 121401 (2003).

\bibitem{pei55}
R. E. Peierls, \textit{Quantum Theory of solids} (Calderon, Oxford
1955) 108.

\bibitem{ong77}
N. P. Ong and P. Monceau, Phys. Rev. B \textbf{16}, 3443 (1977).

\bibitem{aya99}
Ayan Guha, Arindam Ghosh, A. K. Raychaudhuri, S. Parashar, A. R.
Raju and C. N. R. Rao, Appl. Phys. Lett.  \textbf{75}, 3381 (1999).


\bibitem{bra04}
S. Brazovskii and T. Nattermann, Adv. in Phys. \textbf{53}, 177
(2004).

\bibitem{loo02}
P. H. M. van Loosdrecht, B. Beschoten,
            I. Dotsenko and S. van Smaalen, J. Phys. IV France \textbf{12}, 303 (2002).

\bibitem{zaw00}
B. Zawilski, J. Marcus and T. Klein, Europh. Lett. \textbf{50}, 75
(2000).

\bibitem{tes87}
G. X. Tessema and L. Mihaly, Phys. Rev. B \textbf{35}, 7680 (1987).


\bibitem{yam99}
H. Yamada and Y. Ueda, J. Phys. Soc. Jpn. \textbf{68}, 2735 (1999).

\bibitem{yam02}
T. Yamauchi, Y. Ueda and N. Mori , Phys. Rev. Lett. \textbf{89},
057002 (2002).


\bibitem{sie65}
M. J. Sienko  and J. B. Sohn, J. of Chem. Phys. \textbf{44}, 1369
(1965).


\bibitem{yas82}
Yasuyuki Kanai, Seiichi Kagoshima and
             Hiroshi Nagasawa, Jour. Phys. Soc. Jpn. \textbf{51}, 697 (1982).

\bibitem{ued01}
Y. Ueda, H. Yamada, M. Isobe
             and T. Yamauchi , J. Alloys and Compounds  \textbf{317}, 109 (2001).

\bibitem{nag05}
S. Nagai, M. Nishi, K. Kakurai, Y. Oohara, H. Kimura, Y. Noda, T.
Yamauchi, J. I. Yamaura, M. Isobe, Y. Ueda and K. Hirota, J. Phys.
Soc. Jpn. , \textbf{74}, 1297 (2005).

\bibitem{pre03}
C. Presura, M. Popinciuc, P. H. M. van Loosdrecht, D. van der Marel
and M. Mostovoy, Phys. Rev. Lett. \textbf{90}, 026402 (2003).

\bibitem{pres03}
C. Presura , \textit{Energetics and ordering in strongly correlated
oxides as seen in optics}, Ph. D. Thesis, University of Groningen,
The Netherlands, 2003.

\bibitem{samples}
    {The values of the resistivity at a given temperature differ slightly for different samples,
          presumably due to small variations in the sodium stoichiometry \cite{yam99}.}


\bibitem{bra80}
S. A. Brazovskii, Sov. Phys. JETP \textbf{51}, 342 (1980).

\bibitem{field}
    {The electric field is defined as the measured voltage divided by the
          probe contact separation.}

\bibitem{mih88}
G. Mih\`{a}ly, P. Beauchene, J. Marcus,
           J. Dumas and C.Schlenker, Phys. Rev. B \textbf{37}, 1047 (1988).

\bibitem{fle86}
R. M. Fleming, R. J. Cava, L. F. Schneemeyer, E. A. Rieterman and R.
G. Dunn, Jour. Phys. Rev. B. \textbf{33}, 5450 (1986).


\bibitem{wan99}
Dingli Wang, Qingming Xiao, Wufeng Tang,
             Tongyun Zhao, Jing Shi, Decheng Tian and Mingliang Tian, Modern Phys. Lett. B.  \textbf{13}, 109 (1999).

\bibitem{bea86}
P. Beauchene, J. Dumas, A. Janossy, J. Marcus and C. Schlenker,
Physica B \textbf{143}, 126 (1986).

\bibitem{kun05}
S. Yue, C. A. Kuntscher, M. Dressel, S. van Smaalen, F. Ritter and
W. Assmus,  cond-mat/0501332.

\bibitem{sch89}
C. Schlenker, \textit{Low-Dimensional Electronic Properties
            of Molibdenum Bronzes and Oxides} (Kluwer Academic Publishers, 1989).

\bibitem{fro54}
H. Fr\"{o}lich, Proc. R. Soc. \textbf{A 223}, 296 (1954).


\bibitem{oga05}
N. Ogawa, K. Miyano and S. Brazovski, Phys. Rev. B \textbf{71},
075118 (2005).

\bibitem{mon82}
P. Monceau, J.Richard and M. Renard, Jour. Phys. Rev. B.
\textbf{25}, 931 (1982).

\bibitem{gru81}
G. Gr\"uner, A. Zawadowski and P.M. Chaikin, Phys. Rev. Lett.
\textbf{46}, 511 (1981).


\bibitem{bar79}
J. Bardeen , Phys. Rev. Lett. \textbf{42}, 1498 (1979).

\bibitem{lem99}
S. G. Lemay, R. E. Thorne, Y. Li and J. D. Brock, Phys. Rev. Lett.
\textbf{83}, 2793 (1999).


\bibitem{fuk78}
H. Fukuyama and P. A. Lee, Phys. Rev. B. \textbf{17}, 535 (1978).

\bibitem{lee79}
P. A. Lee and T. M. Rice, Phys. Rev. B. \textbf{19}, 3970 (1979).

\bibitem{yama02}
J. I. Yamaura, M. Isobe, H. Yamada, T. Yamauchi and Y. Ueda, J. of
Phys. and Chem. of Solids \textbf{63}, 957 (2002).

\bibitem{and64}
P. W. Anderson and Y. B. Kim, Rev. Mod. Phys. \textbf{36}, 39
(1964).


\bibitem{mot69}
N. F. Mott, Philos. Mag. \textbf{19}, 835 (1969).

\bibitem{mot79}
N. F. Mott and E. A. Davis, \textit{Electronic Processes in
Non-Crystalline Materials} (Clarendon Press, Oxford 1979).



\end{thebibliography}
\end{document}